\documentclass[12pt]{article}
\usepackage{epsfig}

\newcommand{\be}{\begin{equation}}
\newcommand{\ee}{\end{equation}}
\newcommand{\bea}{\begin{eqnarray}}
\newcommand{\eea}{\end{eqnarray}}

\renewcommand{\theequation}{\arabic{equation}}

\begin{document}
\begin{titlepage}

\vspace{1in}

\begin{center}
\Large
{\bf The Ermakov--Pinney Equation in Scalar Field Cosmologies}

\vspace{1in}

\normalsize

\large{Rachael M. Hawkins$^1$ and James E. Lidsey$^2$}

\normalsize
\vspace{.7in}

{\em Astronomy Unit, School of Mathematical 
Sciences,  \\ 
Queen Mary, University of London, 
Mile End Road, LONDON, E1 4NS, U.K.}

\end{center}

\vspace{1in}

\baselineskip=24pt
\begin{abstract}
\noindent It is shown that 
the dynamics of cosmologies 
sourced by a mixture of perfect fluids and self--interacting 
scalar fields are described by the non--linear, Ermakov--Pinney equation. 
The general solution of this equation can be expressed 
in terms of 
particular solutions to a related, 
linear differential equation. 
This characteristic is employed to derive exact cosmologies 
in the inflationary and quintessential scenarios. 
The relevance of the Ermakov--Pinney equation 
to the braneworld scenario is discussed.

\end{abstract}

PACS NUMBERS: 98.80.Cq

\vspace{.7in}
$^1$Electronic mail: R.M.Hawkins@qmul.ac.uk

$^2$Electronic mail: J.E.Lidsey@qmul.ac.uk
\end{titlepage}

%\double 

\setcounter{equation}{0}

\def\theequation{\arabic{equation}}

{\em 1. Introduction}:
Observations of the cosmic microwave background (CMB) 
power spectrum \cite{boomerang} now provide strong 
support for the inflationary 
scenario \cite{early} (for recent reviews, see, e.g., 
\cite{reviews}). The simplest mechanism for inflation utilises the potential 
energy associated with the self--interaction 
of a scalar inflaton field to drive the accelerated expansion.
High redshift observations of type Ia supernovae
suggest
that the universe is experiencing another phase of accelerated expansion
at the present epoch \cite{Ia}. 
This, combined with the 
CMB data, presents a picture of the universe dominated by   
a dark energy component \cite{per,jha}. 
One possible source 
of this dark energy is a scalar 
quintessence field
that interacts with baryonic 
and non--baryonic matter in 
such a way that its potential energy is currently 
dominating the cosmic dynamics \cite{qu}. 
The observations favour a model of structure formation where  
70\% of the energy density 
of the universe is presently in the 
form of quintessence \cite{jha}. 
The remaining fraction of the energy density 
is comprised of visible and cold dark matter  
which collectively act as a pressureless perfect fluid. 

The ekpyrotic scenario
has recently been proposed as an alternative to the standard inflationary 
cosmology \cite{khury}. 
In this scenario the big bang is interpreted as the collision 
of two domain walls or branes travelling through a fifth dimension. 
Before the collision, 
the effective dynamics on the four dimensional branes is
described by Einstein's gravity 
minimally coupled to a 
self--interacting scalar field. 
The field parametrizes 
the separation between the branes
and at early times
slowly rolls down a 
{\em negative} potential. 
This results in an accelerated collapse
of the universe and since the 
field is minimally coupled, 
its energy density is related to the Hubble 
parameter by the standard Friedmann equation. 
Thus, self--interacting scalar fields 
play a central role in modern cosmology and
in view of the above developments, 
it is important to investigate cosmologies that contain 
both a scalar field and a perfect fluid.
 
In this paper, 
we develop an analytical approach 
to models of this type by expressing 
the cosmological field equations in terms of an 
Ermakov system \cite{ermakov,lewis,rr}. In general, 
an Ermakov system is
a pair of coupled, second--order, non--linear 
ordinary differential equations (ODEs) \cite{rr}
and such systems 
often arise in studies  
of nonlinear optics \cite{op}, 
nonlinear elasticity \cite{elas}, 
molecular structures \cite{grc}, quantum field theory in curved 
spaces \cite{finelli} and quantum cosmology \cite{qcos}.
(For further 
references, see, e.g., Refs. \cite{srb}). 

In the one--dimensional case, 
the two equations decouple and
the system reduces to a single equation 
known as the Ermakov--Pinney equation \cite{ermakov,pinney}. 
This is given by\footnote{Eq. (\ref{milnepinney}) 
is sometimes referred to as the Milne--Pinney equation 
\cite{milne}.}
\begin{equation} 
\label{milnepinney}
\frac{d^2 b}{d \tau^2} +Q(\tau ) b =\frac{\lambda}{b^3}  ,
\end{equation} 
where $Q$ is an arbitrary function of $\tau$ and $\lambda$ 
is a constant. 
Although Eq. (\ref{milnepinney}) 
is a non--linear ODE, it exhibits 
a remarkable superposition 
property \cite{lewis,lutzky}
that implies that 
its general solution can be 
expressed directly in terms of particular 
solutions to the related {\em linear}, second--order ODE, where $\lambda =0$
\cite{pinney}.  

{\em 2. The Ermakov--Pinney Equation in Cosmology}: 
We begin by deriving the Ermakov--Pinney 
equation in a cosmological context. The 
field equations 
for a spatially flat, 
Friedmann--Robertson--Walker (FRW) universe 
with a scalar field and perfect fluid 
matter source are given by 
\begin{eqnarray}
\label{friedmann}
H^2 = \frac{\kappa^2}{3} \left( \frac{1}{2} \dot{\phi}^2 +
V(\phi) + \frac{D}{a^n} \right) \\
\label{phieom}
\ddot{\phi} +3H\dot{\phi} +\frac{dV}{d\phi} =0  ,
\end{eqnarray}
where $\rho_{\phi} \equiv \dot{\phi}^2 /2 +V (\phi )$ 
is the energy density of the scalar field with potential 
$V(\phi )$, 
$\rho_{\rm mat} \equiv D a^{-n}$ is the energy density 
of the barotropic perfect fluid with 
equation of state $p_{\rm mat}
= [(n-3)/3]\rho_{\rm mat}$, 
a dot denotes differentiation with respect to cosmic time, 
$t$, $H\equiv \dot{a}/a$ represents the Hubble parameter, 
$a$ is the scale factor of the universe, 
$D$ is an arbitrary, positive constant, $0\le  n \le 6$,
$\kappa^2 \equiv 8\pi m_P^{-2}$ and $m_P$ is the Planck mass. 
Eq. (\ref{phieom}) represents the conservation of the 
energy--momentum of the 
scalar field and can be expressed in the form 
$\dot{\rho}_{\phi} =-3H\dot{\phi}^2$.

One method of reducing this second--order, non--linear 
pair of ODEs (\ref{friedmann})--(\ref{phieom}) 
is to write them as a first--order system. This can be done 
if the scalar field 
varies monotonically with cosmic time.  
It then follows that Eq. (\ref{phieom}) can be expressed 
as \cite{lidsey}
\begin{equation}
\label{mono}
\frac{d\rho_{\phi}}{d\phi} 
=-3H \dot{\phi}
\end{equation}
and the Friedmann 
equation (\ref{friedmann}) then 
takes the form
\begin{equation}
\label{friedmannhat}
\frac{d\chi}{d\phi} \frac{d\rho_{\phi}}{d\phi} 
+\left( n \kappa^2 \right) \chi \rho_{\phi} =-n\kappa^2 D  ,
\end{equation}
where $\chi \equiv a^n$.  
The perfect fluid source leads to a non--trivial right--hand side 
in Eq. (\ref{friedmannhat}).
The general solution to Eq. (\ref{friedmannhat})  
can be expressed in terms of quadratures if the functional 
form of $\rho_{\phi} (\phi )$ is known:
\begin{eqnarray}
a (\phi )= \exp \left[ -\kappa ^2 \int^{\phi} d\tilde{\phi} 
\rho_{\phi} (\tilde{\phi} ) 
\left( \frac{d\rho_{\phi}}{d\tilde{\phi}} \right)^{-1} \right] 
\nonumber \\
\times \left\{ \Pi -n\kappa^2 D \int^{\phi} 
d\tilde{\phi} \left( \frac{d\rho_{\phi}}{d \tilde{\phi}} \right)^{-1}
\exp \left[ n\kappa^2 \int^{\tilde{\phi}} d \varphi
\rho_{\phi} (\varphi )\left( \frac{d\rho_{\phi}}{d\varphi} 
\right)^{-1} \right] 
\right\}^{1/n}   ,
\end{eqnarray}
where $\Pi$ is a constant of integration. The effects of the fluid 
source are contained in the second term.

The disadvantage of this change of variables is that 
it is not valid if the scalar field 
exhibits oscillatory behaviour at any time.
We therefore consider a different 
approach by differentiating 
Eq. (\ref{friedmann}) and substituting 
Eq. (\ref{phieom}). 
This implies that 
\begin{equation}
\label{raych}
\frac{\ddot{a}}{a} -\frac{\dot{a}^2}{a^2} = - 
\frac{\kappa^2}{2} \left( \dot{\phi}^2  + \frac{nD}{3a^n} 
\right)   .
\end{equation}
We now define an effective scale factor, $b$: 
\begin{equation}
\label{newscale}
a \equiv b^{2/n}
\end{equation}
and a new time parameter, $\tau$:
\begin{equation}
\label{newtime}
\frac{d}{dt} \equiv b \frac{d}{d \tau}
\end{equation}
and since $b$ is positive--definite, 
$\tau$ is a monotonic increasing function of 
cosmic time, $t$.
Eq. (\ref{raych}) then transforms 
into the second--order, non--linear Ermakov--Pinney 
equation
\begin{equation}
\label{fullpinney}
\frac{d^{2}b}{d\tau^2}+\frac{n\kappa^2}{4} \left( \frac{d\phi}{d \tau} 
\right)^2 b=-\frac{Dn^2\kappa^2}{12} \frac{1}{b^{3}}
\end{equation} 
and comparison with Eq. (\ref{milnepinney}) then implies that
\begin{equation}
\label{Qexp}
Q = \frac{n\kappa^2}{4} \left( \frac{d\phi}{d \tau} 
\right)^2 , \qquad 
\lambda = -\frac{Dn^2\kappa^2}{12}.
\end{equation}

The change of variables (\ref{newscale}) and 
(\ref{newtime}) is interesting because 
Pinney \cite{pinney} has shown that two linearly independent 
solutions, $\left( b_1 (\tau ) , b_2 (\tau ) \right)$, 
of the time--dependent oscillator
equation
\begin{equation}
\label{helmholtz}
\frac{d^2 b}{d \tau^2} +Q (\tau) b =0
\end{equation}
can be combined to give the general solution to the Ermakov--Pinney 
equation (\ref{milnepinney}): 
\begin{equation}
\label{gensol}
b_P = \left[ A b_1^2 +Bb_2^2 +2C b_1b_2 \right]^{1/2}  ,
\end{equation}
where 
$\{ A,B,C \}$ 
are constants satisfying the constraint
equation 
\begin{equation}
\label{constraint}
AB-C^2 = \frac{\lambda}{W^2}
\end{equation}
and the Wronskian 
\begin{equation}
\label{wronskian}
W \equiv b_1 \frac{db_2}{d\tau} - b_2 \frac{db_1}{d\tau}
\end{equation}
is a constant due to Abel's theorem. 
It can be verified by direct differentiation 
that Eq. (\ref{gensol}) does 
indeed 
satisfy Eq. (\ref{milnepinney}). 
An important property of Eq. (\ref{helmholtz}) is that if a 
particular solution, $b_1 (\tau )$, 
is known, 
a second solution can be written in terms 
of a quadrature: 
\begin{equation}
\label{second}
b_2 (\tau) = W b_1 (\tau) \int^{\tau} 
\frac{d \tilde{\tau}}{b_1^{2} (\tilde{\tau})}  .
\end{equation}
Thus, the general solution to the Ermakov--Pinney 
equation (\ref{fullpinney}) is found in terms of a 
particular solution to Eq. (\ref{helmholtz}). 

The perfect fluid component $(D \ne 0)$
results in the 
non--linear 
sector on the right--hand side of Eq. (\ref{fullpinney}) 
and  
the effects of the scalar field 
are parametrized entirely in terms of the 
function, $Q$,
defined in Eq. (\ref{Qexp}). 
We refer to this function as the `oscillator 
potential'.
It is worth noting that Eq. (\ref{fullpinney}) 
is independent of the specific 
functional form of the scalar field 
interaction potential, $V(\phi )$. 

For the pure scalar field model
$(D=0)$, Eq. (\ref{helmholtz}) 
may also 
be interpreted as a 
classical analogue of the zero--energy Schr\"odinger 
equation, where the scale factor, $b$, plays the 
role of the `wavefunction' and the 
`wave vector', $Q(\tau )$, 
is determined 
by the kinetic energy of the scalar field.
An equation of this form has arisen previously 
in 
studies of FRW cosmologies containing only a perfect fluid 
source with a
constant equation of state $\omega = p_{\rm mat}/\rho_{\rm mat}$
\cite{faraoni} 
and for the special case of a 
self--interacting scalar field 
that satisfies this condition \cite{coop}. 
This analogy may be 
extended further
by considering an alternative reparametrization, 
where a new scale factor and time variable 
are defined by 
$c \equiv a^{-n /2}$ and 
$\sigma \equiv \int dt c$, respectively.
It follows that Eq. (\ref{raych}) 
may then be expressed as a one--dimensional 
Schr\"odinger equation:
\begin{equation}
\label{energy}
\frac{d^2 c}{d \sigma^2} + \left[ 
E - P( \sigma ) \right] c 
=0  ,
\end{equation}
where $E = -(n^2 \kappa^2 D)/12$
represents the total energy of the system, 
and is determined by the momentum of 
the perfect fluid, and 
the potential energy is given by $P= 
(n \kappa^2 /4)(d\phi /d \sigma )^2$ 
in terms of the scalar field. 

There is a close relationship between 
the Schr\"odinger equation (\ref{energy}) and 
the Ermakov--Pinney equation (\ref{milnepinney}), 
as discussed by Milne \cite{milne},  
who developed a method for solving the 
former equation in full generality in terms 
of a particular solution of the latter,
in such a way that 
the oscillatory behaviour of the wavefunction  
is manifest. In this
approach,  
the 
general solution to Eq. (\ref{helmholtz}), 
or equivalently Eq. (\ref{energy}),
is expressed in the form
\begin{equation}
\label{milnesol}
b_H= U \psi (\tau ) \cos \left( \sqrt{\lambda} \theta (\tau ) +
\epsilon \right) ,
\end{equation}
where $\{ U , \epsilon \}$ are 
arbitrary constants and $\{ \psi (\tau ) , 
\theta (\tau ) \}$ are functions 
that need to be determined. Substituting 
this ansatz into the homogeneous equation (\ref{helmholtz}) reveals that 
Eq. (\ref{milnesol}) represents the general 
solution if these functions satisfy the 
coupled system of ODE's:
\begin{eqnarray}
\label{betaeq}
\frac{d^2 \psi}{d \tau^2} +Q(\tau ) \psi = \frac{\lambda}{\psi^3}
\\
\label{thetaeq}
\frac{d\theta}{d\tau} = \frac{1}{\psi^2}
\end{eqnarray}
and it follows, therefore, that the general 
solution to the
Schr\"odinger equation (\ref{helmholtz}) can be 
deduced, at least 
with respect to the quadrature,
$\theta = \int^{\tau} d\tilde{\tau} 
\psi^{-2}(\tilde{\tau} )$,  
if a solution 
to the Ermakov--Pinney equation (\ref{betaeq})
is known.

Furthermore, two linearly independent 
parametric solutions to Eq. (\ref{helmholtz}) 
are given by $b_1= \psi \cos \left( \sqrt{\lambda} 
\theta \right)$ and 
$b_2 = \psi \sin \left( \sqrt{\lambda} 
\theta \right)$, where the Wronskian is  
$W^2=\lambda$, $\theta (\tau )$ satisfies Eq. (\ref{thetaeq})
and $\psi(\tau )$ is a particular solution to Eq. (\ref{betaeq}). 
Substituting these two solutions into 
Eq. (\ref{gensol}) gives 
\begin{equation}
b_P =\psi \left[ A \cos^2 \left( \sqrt{\lambda} \theta 
\right) +B 
\sin^2 \left( \sqrt{\lambda} \theta \right) +(AB-1)^{1/2}
\sin \left( 2\sqrt{\lambda} \theta \right) \right]^{1/2}  ,
\end{equation}
where  we have employed 
Eq. (\ref{constraint}).
This implies that the general 
solution to the Ermakov--Pinney equation (\ref{milnepinney}) 
can be determined in terms of a particular 
solution, $\psi (\tau)$, to this equation 
\cite{lutzky}.

To summarize thus far, the dynamics 
of a pure scalar field 
cosmology is 
determined by a one--dimensional 
oscillator equation with a time--dependent 
frequency. Depending on the 
parametrization chosen, the inclusion of 
a perfect fluid source 
results in an effective constant 
shift in 
the
total energy of the system, as in  Eq. (\ref{energy}),
or equivalently
introduces a non--linear term 
as is the case in Eq. (\ref{fullpinney}). 
In particular,
the problem of solving Eq. (\ref{fullpinney}) 
for a coupled perfect fluid and scalar field cosmology 
has been reduced to solving Eq. (\ref{helmholtz}) 
for a given functional form of $Q (\tau )$.
This is interesting because this latter equation 
has been studied extensively in the literature. 
For example, 
algorithms that extend 
the Wentzel, Kramers and Brilluen (WKB) 
technique have recently been developed 
to find power series solutions to  
Eq. (\ref{helmholtz})
\cite{maltzev}. For related work, see also Ref. \cite{bogdan}. 

{\em 3. Algorithms for Solving the Field Equations}:
We now proceed to 
outline an algorithm for 
deriving exact cosmological models when 
both a scalar field and perfect fluid are 
dynamically significant. 
Attention to date has primarily focused
on pure scalar field models 
\cite{lidsey,lm,em,others}. Some specific models with a perfect fluid 
were also recently analyzed 
\cite{rb}.
>From a particle physics perspective the potential 
of the scalar field is the fundamental quantity 
that defines the model. In general,  however, 
Eqs. (\ref{friedmann}) and (\ref{phieom}) are 
very difficult, if not impossible, to solve unless the potential 
has a specific form, as in the exponential case \cite{lm}.
There are several possible routes to take when solving 
the Ermakov-Pinney equation (\ref{milnepinney}).
Given its properties as discussed above, one possibility is to 
specify the time dependence of the scalar field, $\phi =\phi (\tau )$, 
since the functional form of $Q=Q(\tau )$ is 
unaltered when deriving the solution (\ref{gensol}) from 
the corresponding solutions to the homogeneous equation (\ref{helmholtz}).

>From the cosmological point of view, however, 
it is the time--dependence of the scale factor that is important.
In particular, the effective equation of state and the 
potential of the scalar field can in principle be determined   
directly 
from high redshift observations if the dependence 
of the scale factor on redshift is known to a 
sufficiently high accuracy 
\cite{starpot,ge}. This question has recently attracted renewed 
interest in light of the proposed Supernova/Acceleration Probe 
(SNAP) \cite{ge,snap}.

Instead of choosing the potential, therefore, 
an alternative approach 
is to invert the problem 
by first specifying the 
dependence
of the scale factor on cosmic time \cite{lm,em}. 
We may begin with an {\em ansatz} for the scale factor, $a_1 (t)$, 
or equivalently, the rescaled function, $b_1 (t)$, 
and use 
Eq. (\ref{helmholtz}) to find $Q(\tau)$. Integrating this 
function then yields the time--dependence of the 
scalar field. The general 
solution to the Ermakov--Pinney 
equation can be deduced immediately from 
Eqs. (\ref{gensol}) and (\ref{second}). 
The scale factor for the Ermakov-Pinney equation in parametric form 
then follows from 
Eq. (\ref{newscale}). Its dependence on cosmic time 
may be evaluated by integrating 
Eq. (\ref{newtime}) and inverting the result. 
Finally, the scalar field 
potential is reconstructed 
directly from the Friedmann equation (\ref{friedmann}), 
rewritten in terms of the new variables  
(\ref{newscale}) and (\ref{newtime}):
\begin{equation}
\label{pottau}
V(\tau ) =\frac{12}{n^2 \kappa^2} \left( 
\frac{db_P}{d \tau} \right)^2 -\frac{1}{2} b^2_P 
\left( \frac{d \phi}{d \tau} \right)^2  
-\frac{D}{b_P^2}  .
\end{equation}
The form of $V(\phi )$ then follows by inverting 
$\phi (\tau )$ and substituting $\tau(\phi)$ into Eq. (\ref{pottau}).

The various possible algorithms are illustrated in Figure 1. 

\begin{figure}[ht]
\centerline{\epsfig{figure=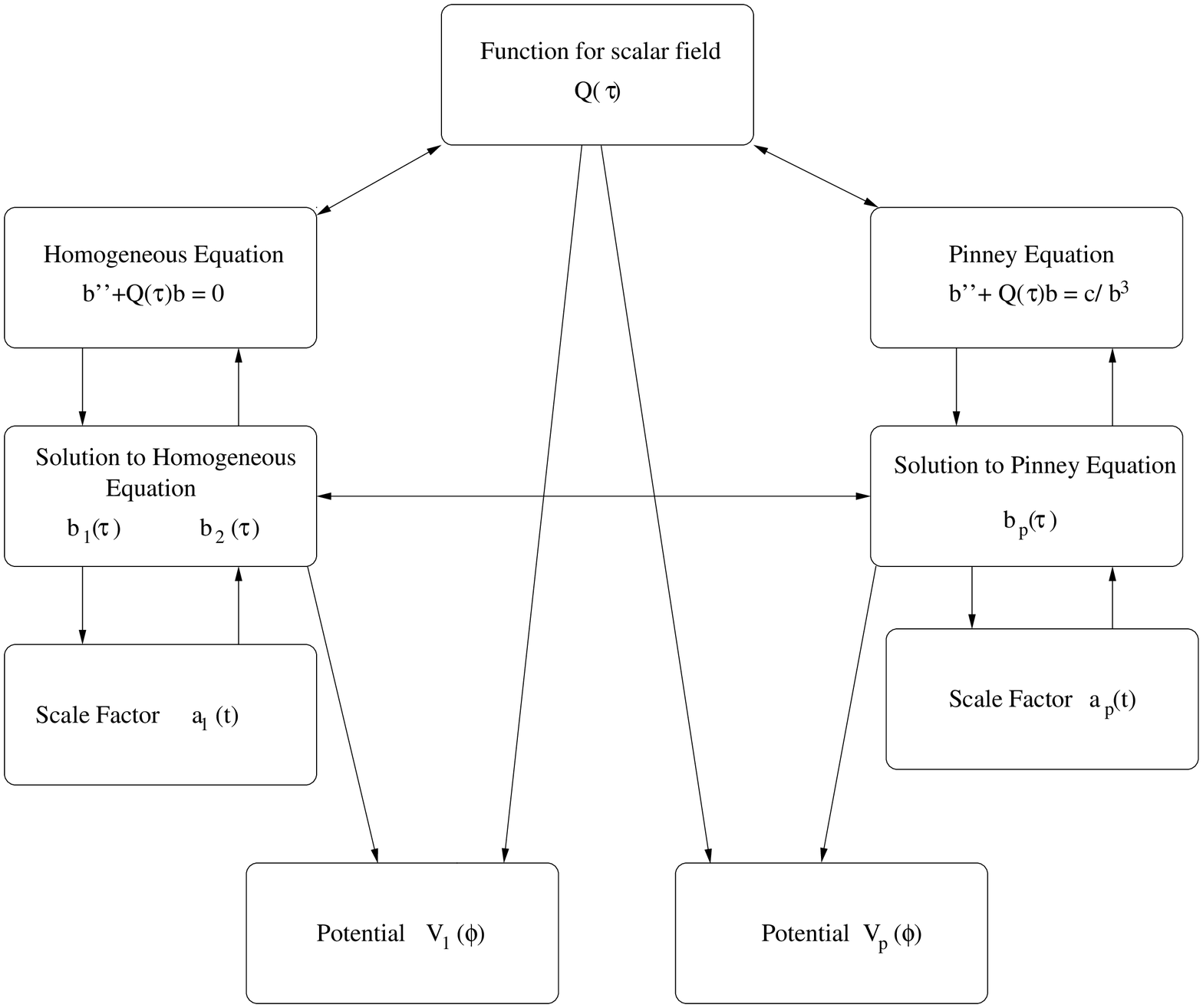,height=100mm}}
\caption{\small Illustrating 
the possible routes for finding 
the scalar field potential in a pure scalar field system and 
in a 
system comprised of both a scalar 
field and a 
perfect fluid. The cosmological Ermakov-Pinney equation
(\ref{fullpinney}) plays a central role.
The two potentials can be found from a given functional form of the 
oscillator potential, $Q(\tau)$, or from the scale factors, $a(t)$.
In principle, it is consistent to begin at any point on this Figure. From  
a cosmological perspective, the scale factor is the 
primary function of interest, and specifying this parameter 
from the outset is physically well motivated.}
\end{figure}

{\em 4. Exact Cosmologies Sourced by Scalar Fields and 
Perfect Fluids}:
We now illustrate this procedure with a number 
of examples. The simplest case is that of the scale factor as an 
exponential function of time $a_{1}(t)=e^{2t/n}$ where the power has been 
chosen so that $b_{1}(\tau)=\tau$ and therefore $b_{2}(\tau)=1$. 
Such a choice produces a static scalar field, $d\phi /d \tau =0$, 
corresponding to a pure cosmological 
constant. The solution for a cosmological constant with a perfect 
fluid is already known \cite{harrison}, but this example illustrates 
the generality of the technique. Thus, 
from Eqs. (\ref{newscale}) and (\ref{gensol}), the general 
solution for the scale factor when a perfect fluid is 
present is given by 
\begin{equation}
\label{perfcos}
a_P = \left( A \tau^2 +2C \tau +B \right)^{1/n}  ,
\end{equation}
where $AB-C^2 =-Dn^2\kappa^2/12$. Since the scalar 
field remains independent of time, the potential is 
also a constant, $V= 12A/(\kappa^2 n^2)$, from 
Eq. (\ref{pottau}). Finally, 
the dependence of the scale factor on cosmic 
time is deduced by using Eq. (\ref{newscale}) and integrating Eq. 
(\ref{newtime}): 
\begin{equation}
\label{scaleconstant}
a_P= \left( \frac{-\lambda}{A} \right)^{1/n}
{\rm sinh}^{2/n} \left( \sqrt{A} t \right)  .
\end{equation} 

A more general {\em ansatz} for the scale factor is given by 
\begin{equation}
\label{aplusminus}
a_{\pm}(t) =  G t^{q_{\pm}} ,
\end{equation}
where $q_{\pm}$ are constants and $G
\equiv [2/(2+nq_{\pm} ) ]^{q_{\pm}}$.
Using the relationship between $a$ and $b$ given in Eq. (\ref{newscale}), 
it is possible to express the two solutions to the homogeneous equation as
\begin{equation}
\label{bplusminus}
b_{\pm} = \tau^{p_{\pm}} , 
\end{equation}
where $p_{\pm}= nq_{\pm}/(2+nq_{\pm})$. 
Solving the homogeneous equation Eq. 
(\ref {helmholtz}) then gives the form of the scalar field $\phi(\tau)$,
\begin{equation}
\label{powerscalar}
\phi =F \ln \tau , 
\end{equation}
where the constant $F$ is related to the power $p_{\pm}$ 
via the quadratic equation
\begin{equation}
\label{pandf}
\qquad p_{\pm} = 
\frac{1\pm \sqrt{1-n\kappa^2 F^2}}{2}.
\end{equation}

The Wronskian is  
$W^2 =1- n\kappa^2 F^2$ and, 
for consistency, we assume that $n\kappa^2 F^2 < 1$ 
in what follows. The 
general solution to Eq. (\ref{fullpinney}) is then found by substituting 
Eq. (\ref{bplusminus}) into 
Eq. (\ref{gensol}) and is
\begin{equation}
\label{powergeneral}
a_P = \left( A\tau^{2p_+} +B \tau^{2p_-} +2C \tau 
\right)^{1/n}  .
\end{equation}

The analytical form of the potential 
deduced from Eq. (\ref{pottau}) 
is a combination of exponential functions of the 
scalar field. The simplest case 
arises for  
$A=B=0$. Eqs. (\ref{Qexp}) and (\ref{constraint})
then imply a relationship between the constants:
\begin{equation}
\label{related}
D= \frac{12C^2}{n^2\kappa^2} \left( 1-n\kappa^2 F^2 \right)
\end{equation}
and consequently, 
the scale factor, the scalar field and its potential are given by 
\begin{eqnarray}
\label{attractor}
a_P (t) = \left( C t \right)^{2/n} \nonumber \\
\phi (t) = F\ln \left( \frac{C}{2} \right) + 2F \ln t \nonumber 
\\
V (\phi ) = \frac{CF^2 (6-n)}{n} \exp \left( -\frac{\phi}{F} 
\right).
\end{eqnarray} 

The solution (\ref{attractor}) describes a cosmology where 
the energy density of the scalar field tracks 
that of the 
perfect fluid 
in a way that leaves 
the dependence of the scale factor 
on cosmic time unaltered 
from the pure perfect fluid model \cite{wetterich}. 
The self--interaction coupling, $F$, 
of the scalar field is constrained only 
by the condition that the 
fluid integration 
constant, $D$, 
be real.  

Eq. (\ref{attractor}) 
represents a past or future attractor 
for the more general classes of solutions 
given by Eq.
(\ref{powergeneral}). For example, 
if $A \ne 0$ and $B=0$ 
the potential (shown in Figure 2 for the case 
of a pressureless fluid) is given by 
\begin{equation}
\label{Apot}
V= \frac{1}{Y} \left[ \frac{12}{n^2\kappa^2} 
Z^2 - \frac{F^2Y^2}{2} e^{-2\phi /F} -D \right]   ,
\end{equation}
where
\begin{eqnarray}
Y \equiv Ae^{2p_+ \phi /F} +2C e^{\phi /F} 
\nonumber \\
Z \equiv Ap_+ e^{[(2p_+-1)\phi /F]} +C.
\end{eqnarray}

\begin{figure}[ht]
\centerline{\epsfig{figure=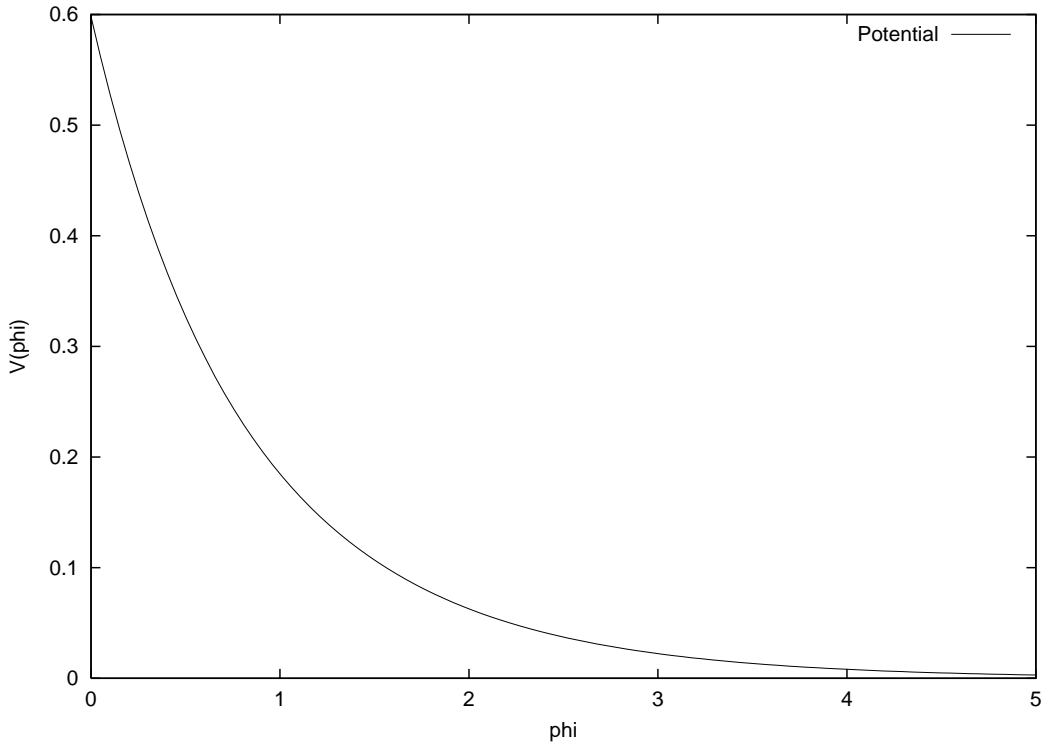,height=100mm}}
\caption{\small 
The potential (\ref{Apot}) 
for $A =1$, $B=0$, $C=\sqrt{3}$, $D=1$, 
$F=1/2$, $\kappa =1$ and $n=3$. The potential asymptotes 
to an exponential form at high and low energies.}  
\end{figure}

In the early--time limit $(\tau \rightarrow 0 )$, 
Eq. (\ref{Apot}) asymptotes to Eq. (\ref{attractor}).
On the other hand, 
the potential reduces to 
\begin{equation}
\label{potinf}
V_{\infty} =
\frac{A}{2} \left( \frac{24p_+^2}{n^2\kappa^2} -F^2 
\right)  \exp \left[ -\frac{2p_-}{F} \phi \right]
\end{equation}
at late times $( \tau \rightarrow \infty )$,
since $p_+ >1/2$. 
In this limit, the scale factor asymptotes to 
$a \propto t^p$, where $p \equiv 
2p_+/(np_-)$ and it is straightforward 
to verify that this power law behaviour 
is the late--time attractor 
for a universe dominated by a single scalar field with 
an exponential potential, $V \propto 
\exp \left( \sqrt{2}\kappa \phi /\sqrt{p}
\right)$.  
Since $p_+ >p_-$, 
the power 
of the expansion in this limit, $p$, exceeds that of the pure perfect 
fluid model $( a \propto t^{2/n} )$ and 
it is greater than unity if $p_+ >n\kappa F /\sqrt{8}$. 
In this region of 
parameter space, therefore, 
the solution given by Eqs. (\ref{powergeneral}) 
and (\ref{Apot}) describes  a universe that 
behaves as a perfect fluid cosmology at 
early times. However,  
the perfect fluid becomes negligible as the expansion 
proceeds and 
the universe subsequently enters a phase of 
power law,  inflationary expansion 
driven by the exponential potential of a scalar 
field. For $n=3$, the fluid is pressureless and 
current observations indicate that the 
universe is undergoing such a transition at the present 
epoch \cite{Ia,per,jha}. Thus, it has been demonstrated in 
the example outlined here that the algorithm for solving 
the cosmological field equations (\ref{friedmann}) 
and (\ref{phieom}) using the Ermakov-Pinney equation 
(\ref{fullpinney}) has 
generated a new exact solution describing the quintessence scenario. 
Moreover,  exponential potentials often arise in  
superstring--inspired models through non--perturbative effects
\cite{kaloper}. 

Another class of solvable models arises when 
the scale factor is a trigonometric function of $\tau$. 
The two solutions to Eq. (\ref{helmholtz}) 
are $b_1 =\sin \gamma \tau$ and 
$b_2 = \cos \gamma \tau$, where $\gamma$ is a constant 
and the Wronskian is $W=-\gamma$.
Given this ansatz for the scale factor, 
the scalar field varies linearly with $\tau$, such that
\begin{equation}
\label{trigphi} 
\phi =\frac{2 \gamma}{\sqrt{n} \kappa} \tau  .
\end{equation}
The solution to Eq. 
(\ref{fullpinney}) is then given by 
\begin{equation}
\label{trigpinney}
a_P = \left[ A \sin^2 \gamma \tau +B \cos^2 \gamma \tau 
+C \sin 2 \gamma \tau \right]^{1/n}  .
\end{equation}
In general, 
the potential is a complicated 
expression of trigonometric functions. However, 
for the particular choice $A=B=0$, 
Eq. (\ref{pottau}) 
takes the simple form
\begin{equation}
V=-\frac{2C\gamma^2}{n^2\kappa^2} (n+6) \sin \left( \sqrt{n} \kappa 
\phi \right) 
\end{equation}
and it follows from 
Eq. (\ref{trigpinney}) that $V \propto a_P^n$. 
Consequently, the 
potential energy of the scalar field
is inversely proportional to the energy
density of the perfect fluid. 

So far in this section it has been demonstrated 
that once a solution to the pure scalar field system 
given in Eq. (\ref{helmholtz}) is known, a solution to 
the Ermakov-Pinney equation (\ref{fullpinney}) 
describing a scalar field 
and perfect fluid cosmology can be found. Before concluding 
this section, 
we employ the correspondence 
summarized in Eqs. (\ref{milnesol})--(\ref{thetaeq})
to show the reverse procedure, i.e., to derive a pure scalar field 
cosmology from a known solution 
to the Ermakov--Pinney equation (\ref{betaeq}). 
We invoke the ansatz 
$\psi(t)=\Lambda e^{\Lambda t}$ which in terms of the rescaled time 
variable $\tau$ is $\psi = \Lambda \tau$, where 
$\Lambda$ is a constant.
Substituting this ansatz into Eq. 
(\ref{betaeq}) and solving Eq. (\ref{Qexp}) for the scalar 
field gives
\begin{equation}
\label{phiexp}
\phi = \mp \left( \frac{4\lambda}{n\kappa^2} 
\right)^{1/2} \frac{1}{\Lambda^2 \tau}  ,
\end{equation}
where the integration constant is specified 
to be zero without loss of generality.
Integrating Eq. (\ref{thetaeq}) 
implies that $\theta = -1/(\Lambda^2 \tau )$
and after specifying $U=1$ and $\epsilon =0$ for simplicity,  
the scale factor of the universe 
and the potential for such a pure scalar field model are 
given by
\begin{equation}
\label{scalecos}
a=\left[ \Lambda \tau \cos \left( 
\frac{\sqrt{\lambda}}{\Lambda^2 \tau} \right) \right]^{2/n}
\end{equation}

\begin{equation}
\label{potcos}
V= \frac{2\Lambda^2}{n^2\kappa^2} \left[ 6 \left( 
\cos \varphi +\varphi \sin \varphi \right)^2 -n \varphi^2 
\cos^2 \varphi \right]   ,
\end{equation}
where $\varphi \equiv \sqrt{n}\kappa \phi /2$.
In the small field limit, $\varphi \ll 1$, 
the potential (\ref{potcos}) approximates 
to an inverted harmonic 
oscillator and for sufficiently small 
values of the scalar field, the universe 
undergoes inflation \cite{bine}. 
Potentials of this form generically 
produce a spectrum of 
density perturbations in the post--inflationary 
universe that increases in amplitude on larger scales 
without producing a detectable signal 
of gravitational waves \cite{natural}. 
Observations of CMB anisotropies presently favour such  a 
spectrum \cite{tegmark}.  

{\em 5. Braneworld Scenarios}:
It is also of interest to consider whether
the approach outlined above can be applied to other cosmological scenarios.
Considerable attention 
has been focused recently on the 
possibility 
that the 
observable universe may be 
viewed as a domain 
wall or brane \cite{brane} that 
is moving along
a timelike geodesic 
in a five--dimensional, 
static, bulk spacetime \cite{rsbrane,othermoves,reall,bv},
where the brane equations of motion are determined by 
the Israel junction conditions \cite{israel}.
An observer confined to the domain wall 
interprets such motion 
in terms of cosmic expansion or contraction.
The junction conditions
relate the second fundamental form (extrinsic curvature) of the 
induced metric on the brane to the energy--momentum tensor 
of matter confined on the brane. 
For the case of a spatially flat brane moving through a 
Schwarzschild--anti de Sitter space,
the effective Friedmann equation on the brane 
is, after some simplifying assumptions 
\cite{othermoves,bv,cline,cline1}, given by
\begin{equation}
\label{friedmannbrane}
H^2 = \frac{\kappa^2}{3} \rho \left( 1+ \frac{\rho}{2\lambda} 
\right) + \frac{\mu_n}{a^n}  ,
\end{equation}
where $\lambda$ is the brane tension, 
$\mu_n$ is related to the mass 
of the black hole in the bulk and $n=4$. 
The standard, linear dependence of the Friedmann equation 
on the energy density, $\rho$, is recovered in the 
low--energy regime, $\rho \ll \lambda$.
Further modifications to the Friedmann 
equation (\ref{friedmannbrane}) may also arise in more general braneworld
settings. For example, a term of the form 
$\Sigma_6 = \mu_6 a^{-6}$, 
where $\mu_6 <0$, arises if 
the black hole carries an electric charge \cite{bv,csaki}.

For this model, the energy--momentum 
tensor of matter on the brane is covariantly 
conserved, and consequently, 
Eq. (\ref{phieom}) must also be satisfied 
for the case of a 
single, self--interacting scalar field. 
Eqs. 
(\ref{phieom}) and (\ref{friedmannbrane})
are therefore relevant to a number of 
braneworld scenarios. 
By differentiating Eq. (\ref{friedmannbrane})
with respect to cosmic time, 
substituting for 
the scalar field from Eq. (\ref{phieom}),  
and using the change of variables 
(\ref{newscale}) and (\ref{newtime}), 
it can be shown that the dynamics of this system 
can be written in the Ermakov-Pinney form
\begin{equation}
\label{epbrane}
\frac{d^2 b}{d \tau^2} + \tilde{Q} b = -\frac{n^2 \mu_n}{4b^3}  ,
\end{equation}
where 
\begin{equation}
\label{tildeQ}
\tilde{Q} \equiv 
\frac{n \kappa^2}{4} \left( 1 + \frac{\rho}{\lambda}
\right) \left( \frac{d\phi}{d \tau}  \right)^2  .
\end{equation}
In Eq. (\ref{epbrane}), the effects that 
lead to     
the quadratic 
dependence of the Friedmann equation 
(\ref{friedmannbrane}) 
on the energy density of the scalar field 
modify only  
the oscillator potential (\ref{tildeQ}),  
whereas the non--linear term 
on the right--hand side 
arises through the bulk black hole contribution.  
Thus, in principle, a similar 
approach to that outlined above in Section 3 may be developed for 
analytically investigating the importance of these terms 
on braneworld cosmologies.
For example, if a given 
solution is known for the case where 
there is no black hole in the bulk \cite{hawkins}, 
it may serve as a seed for generating a 
corresponding cosmology when the black hole is 
present. This is interesting since it has recently 
been shown that such terms can significantly 
alter the qualitative dynamics of 
braneworld cosmologies \cite{mm}.  

{\em 6. Summary}: 
In this paper it has been demonstrated that 
after a suitable redefinition of 
variables, Eqs. (\ref{friedmann})--(\ref{phieom}) 
can be related to the simplest Ermakov system, 
thus reducing the problem of solving these 
equations to solving a single second--order, linear ODE. 
Moreover, 
the nature of the 
Ermakov--Pinney equation 
implies that there 
exists a correspondence 
between a spatially flat 
FRW universe containing a scalar field 
and a cosmology containing both a scalar field and 
a perfect fluid. 
The functional form of the 
scalar field, $\phi (\tau )$, is identical for 
the two models, but 
the potentials, $V(\phi)$, are different
in each case. 
We have proposed an algorithm 
for finding exact solutions for 
a scalar field and perfect fluid model 
by employing a pure scalar field 
cosmology as a seed. The algorithm 
is founded on the property that a particular 
solution to the homogeneous 
equation (\ref{helmholtz}) leads to the general 
solution of Eq. (\ref{milnepinney}). 
It has also been shown that a model containing both a 
scalar field and perfect fluid can 
act as a seed for finding more general pure scalar field 
cosmologies, since a particular solution 
to Eq. (\ref{milnepinney}) allows the general 
solution of Eq. (\ref{helmholtz})
to be deduced \cite{milne}.  
Exact solutions are important because they 
allow one to gain insight into the generic nature 
of cosmologies of this type 
and they also provide a framework for classifying the different 
behaviours that may arise. 

The emphasis of the present 
work has focused on the effect of 
introducing a perfect fluid source. 
However,  
the approach we have developed may 
also  be
employed to consider pure scalar field cosmologies in 
more general spacetimes.
In particular, 
the last term on the right--hand 
side of Eq. (\ref{friedmann}) may  
be interpreted as arising from the spatial curvature of the 
universe if $n=2$, where $D>0$ $(D<0)$ for negative (positive) curvature. 
On the other hand, the shear  
of the spatially flat, anisotropic Bianchi type I metric
leads to a term where $n=6$. 

Finally, it would be of interest to investigate 
whether other classes of scalar field cosmologies can be analyzed 
in terms of an Ermakov system. Scalar--tensor 
cosmologies provide one such example,
where the scalar field plays the 
role of the gravitational coupling \cite{st}. Many 
higher--order and higher--dimensional theories of gravity, 
including the string effective actions, 
may be expressed in a scalar--tensor form \cite{lwc}. 
Since these theories  
are 
conformally equivalent to Einstein gravity that is 
minimally coupled to a scalar 
field, it is to be expected that a similar approach can be 
employed. A related, but distinct, class of theories has recently 
been developed to allow for a
possible variation in the fine--structure constant \cite{m}. 
Observational evidence for such 
an effect has been growing \cite{webb}
and this behaviour can again be parametrized 
in terms of a scalar field that is coupled 
to the matter fields in an appropriate fashion. 

In conclusion, therefore, Ermakov systems have applications 
in many branches of mathematics and physics. 
In this paper, we have found 
that the simplest 
such system is relevant to   
scalar field cosmological models that 
are favoured 
by astrophysical observations. 

\vspace{.3in}
\centerline{\bf Acknowledgments}
\vspace{.3in}

RMH is supported by the Particle Physics 
and Astronomy Research Council (PPARC) and 
JEL is supported by the Royal Society. 
We thank A. Garcia, B. Mielnik, H. Rosu and C. Terrero for 
helpful discussions. JEL thanks the group at the Centro de 
Investigacion y de Estudios Avanzados del IPN, Mexico, 
for hospitality under CONACyT grant 38495-E.

\end{document}